# Use of Mobile Devices in the Classroom to Increase Motivation and Participation of Engineering University Students

C. Guerrero, A. Jaume, C. Juiz, *Senior Member, IEEE* and I. Lera

*Abstract*—The aim of this study was to see whether student participation increased when mobile devices were used in the classroom. We measured the amount of student participative actions when the Socrative tool was used and when it was not used. Our experiment involved a total of 192 students, corresponding to 4 different subjects of Computer Engineering at the Universitat de les Illes Balears, during 2012/2013 and 2013/2014 courses. An independent paired t-test was performed on the measurements. The analysis results show that student participation increases with the use of mobile devices for theory classes and students are willing to participate in class activities and share their own results.

## I. INTRODUCCIÓN

EL uso de teléfonos inteligentes, computadoras portátiles y tablets es habitual entre los estudiantes universitarios, y entre los jóvenes en general. Según Blackboard [1], en Estados Unidos el 98% de los estudiantes universitarios disponen de alguno de estos dispositivos. No es necesario recurrir a un estudio para ser consciente de un hecho tan observable en nuestras propias aulas, pues la mayoría de los alumnos tienen un dispositivo móvil al lado de los cuadernos y libros.

Otro hecho observable, y el cual sufrimos, es la escasa participación del alumnado en las actividades de clase. No transmiten sus dudas ni responden las preguntas que el profesor formula públicamente. Gracias a este diálogo con los alumnos, el profesor puede adaptar las sesiones. Pero por la baja cantidad de participaciones, la adaptación se basa en una pequeña minoría. La creencia habitual es que los alumnos, especialmente en cursos iniciales, no participan en clase por timidez o por estar cohibidos delante del profesor o de sus compañeros.

El profesor puede controlar el ritmo y la calidad del aprendizaje de los estudiantes mediante los comentarios constructivos. Por este motivo una de las responsabilidades del profesorado es detectar y mejorar la falta de participación de los alumnos en el aula.

Para disponer de un feedback rápido, se pueden utilizar sistemas de respuesta de audiencia (ARS), más conocidos como pulsadores (*clickers*). Gracias a su uso se puede evaluar las actividades entre los alumnos, el grado de concentración, la interacción entre ellos y el docente, y la resolución de problemas. Entre sus desventajas cabe destacar el alto coste y la falta de inmediatez en su despliegue y uso.

Los teléfonos inteligentes permiten ejecutar aplicaciones que emulan el uso de estos *clickers*, lo que permite una reducción de costes y una simplificación de la logística. Además, estas aplicaciones son más versátiles y ofrecen más funcionalidades como es, por ejemplo, el envío de mensajes al profesor.

En este trabajo, se usaron los teléfonos móviles de los alumnos con el objetivo de promover y aumentar la participación activa durante las clases. De esta forma, el profesor puede obtener un feedback más completo. Nuestra hipótesis de trabajo era que el uso de herramientas móviles podía mejorar la participación de los alumnos durante las clases y, en consecuencia, incrementar el número de alumnos que realizaban los ejercicios.

El artículo está organizado de la siguiente forma: en la Sección II se expone el estado del arte; en la Sección III se detallan el diseño de los experimentos y los instrumentos utilizados; la Sección IV muestra los resultados obtenidos; la discusión de los resultados se lleva a cabo en la Sección V. Finalmente, en la Sección VI se muestran las conclusiones del trabajo y planteamos alternativas de trabajo futuro.

## II. ESTADO DEL ARTE

Los trabajos analizados demuestran la viabilidad del uso de dispositivos de interacción para detectar, reforzar y mejorar tanto los conceptos teóricos y prácticos como el ambiente

C. Guerrero was with Departament de Matemàtiques i Informàtica, Universitat de les Illes Balears, 07122 Palma, Spain. carlos.guerrero@uib.es
A. Jaume was with Departament de Matemàtiques i Informàtica, Universitat de les Illes Balears, 07122 Palma, Spain. antoni.jaume@uib.es
C. Juiz was with Departament de Matemàtiques i Informàtica, Universitat de les Illes Balears, 07122 Palma, Spain. cjuiz@uib.es
I. Lera was with Departament de Matemàtiques i Informàtica, Universitat de les Illes Balears, 07122 Palma, Spain. isaac.lera@uib.es

positivo en la clase [2].

Sun et al. [3] analizaron el grado de atención del alumnado mediante la comparativa de dos modelos de obtención de respuesta: pulsadores y sistemas de votación por web. En su estudio se establecieron dos grupos (95 y 114 estudiantes respectivamente) y se evaluó la atención emocional, cognitiva y total respecto a tres tipos de cuestiones. Determinaron que ambos sistemas eran totalmente eficaces para captar la atención y generar motivación.

Bojinova y Oigara [4] cuestionaron el uso de pulsadores como herramientas positivas para el aprendizaje. Evaluaron dos cursos durante un semestre en sesiones prácticas, mediante la comparación de resultados con grupos de control. En sus resultados apenas encontraron diferencias entre ambos, aún así destacaron la menor desviación de notas en la clase de tratamiento. Además, mediante un cuestionario directo encontraron que el grupo presentaba un mayor grado de motivación.

Caldwell [5] afirmó que los pulsadores eran herramientas flexibles para la docencia en cualquier nivel educativo y ofrecían una serie de pautas para su uso. Además, su uso incrementaba el grado de atención y retención de conceptos. Su estudio se centró en la descripción de actividades gracias al uso de estos dispositivos. Durante sus clases los efectos fueron predominantemente positivos, en el peor de los casos siendo neutrales. Observó una mayor efectividad en actividades de aprendizaje colaborativo a lo largo de un año en su asignatura de biología con 250 estudiantes.

Álvarez y Llosa [6] combinaron métodos expositivos y mandos específicos para evaluar preguntas directas. Esto les sirvió para detectar y reforzar las debilidades del alumnado. Utilizaron mandos específicos por radiofrecuencia con un coste de 100 € cada uno. Encontraron una pequeña correlación entre el uso de los dispositivos y las notas extraídas.

En el estudio de Simelane y Miji [7] utilizaron los pulsadores como herramientas para validar una estrategia docente centrada en la retención de la atención. Los definieron como herramientas flexibles para obtener una opinión rápida del alumnado. Los pulsadores eran específicos y fueron alquilados durante los seis meses que duró la prueba.

El análisis realizado por Bunce et al [8] cuestionó si el uso de pulsadores disminuía el grado de atención. En su trabajo analizaron los lapsus existentes en cualquier clase y los compararon con las interrupciones creadas por el uso de pulsadores. Mediante el intercalado de preguntas y exposiciones registraron estos periodos de tiempo usando tabletas digitales. Concluyeron que en contra de lo esperado no había una atención continua durante los primeros 10-20 minutos. Desde el inicio de la clase, se producían pequeños ciclos de atención y desatención, alrededor de 4,5 min. Este periodo variaba según el profesor y la materia, pero en ningún caso su efecto era transcendente. Con la utilización de pulsadores, se produjo un aumento de la atención durante la exposición de la pregunta pero su efecto decayó y su falta de atención se prolongó para el siguiente periodo. Concluyeron su trabajo indicando que lo importante era el cambio de metodologías y la alternancia de formatos para evitar la desatención.

## III. MATERIALES Y MÉTODOS

Con el objetivo de incrementar la participación mediante el uso de dispositivos móviles durante las clases, utilizamos la aplicación multiplataforma *Socrative* (www.socrative.com). Se puede utilizar tanto en entorno web como en aplicación móvil, y emula un pulsador tradicional pero con ciertas funcionalidades adicionales como el envío de mensajes de texto al profesor. La aplicación consta de dos módulos, uno para los profesores y otro para los alumnos (Fig. 1).

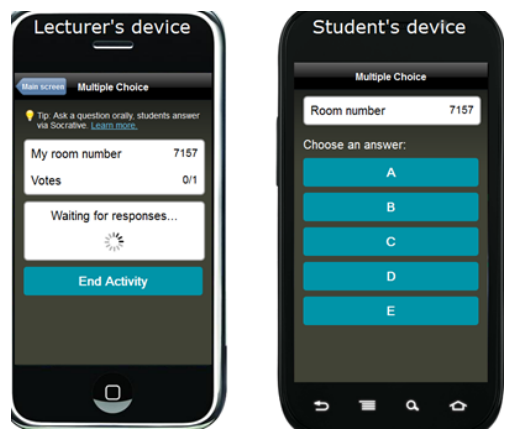

Figura 1. Captura del módulo para los profesores (a la izquierda) y del módulo para los alumnos (a la derecha).

La aplicación permitía al profesorado enviar a los *smartphones* de los alumnos preguntas en tiempo real, la agregación inmediata de resultados y su visualización. El profesor enunciaba una pregunta en clase, y a través de su módulo activaba el tipo de respuesta que deseaba: de selección múltiple, de verdadero o falso, o de texto libre.

Una vez que todos los alumnos habían respondido el profesor recibía un resumen de las respuestas. De esta forma, el profesor podía obtener en tiempo real información del aprendizaje y del conocimiento de los alumnos, y decidir como continuar con la explicación.

Por su parte, los alumnos, al empezar una clase introducían el código que les indicaba el profesor en su módulo. Cuando el profesor activaba una pregunta, recibían una notificación para responder.

Diseñamos un experimento en el que se incluyó el uso de la herramienta *Socrative* con el objetivo de contrastar la hipótesis de que el uso de pulsadores podía aumentar la participación de los alumnos durante las clases en el aula.

*A. Participantes*

En el experimento participaron un total de 192 alumnos, en 4 asignaturas del Grado en Ingeniería Informática de la Universitat de les Illes Balears, durante los cursos 2012/2013 y 2013/2014. En concreto, 36 alumnos pertenecían a la asignatura *Evaluación del Comportamiento de Sistemas Informáticos* ($A_1$), 66 alumnos pertenecían a la asignatura *Arquitectura de Computadores* ($A_2$), 62 alumnos pertenecían a la asignatura de *Programación Avanzada* ($A_3$), y 28 alumnos pertenecían a la asignatura de *Visión por Computador* ($A_4$).

*B. Procedimiento*

El experimento se desarrolló durante una sesión, a final de semestre, de cada una de las asignaturas que se incluyeron en el experimento. Los profesores dividieron cada sesión en bloques de explicación de contenidos ($BC_n$) y en bloques de resolución de ejercicios ($RE_n$).

En los bloques de explicación de contenidos el profesor explicaba el contenido del bloque temático y, durante la explicación, los alumnos podían realizar preguntas. El profesor se encargaba de cuantificar el número de preguntas que le hacían durante un bloque de explicación de contenidos ($P_n$).

En los bloques de resolución de ejercicios, el profesor proponía a los alumnos un ejercicio, relacionado con los contenidos explicados en el bloque de explicación anterior y les daba un tiempo limitado para resolverlo. Al finalizar el tiempo, el profesor recogía las respuestas de los alumnos y cuantificaba el número de respuestas ($R_n$). A continuación explicaba la solución y respondía a las posibles preguntas que realizaban los alumnos a la solución de este ejercicio, cuantificando el número de estas preguntas ($D_n$).

Las sesiones de clase en las que se realizaron los experimentos, se organizaron de forma que se empezaba con un bloque de explicación de contenidos ($BC_a$), seguido de dos bloques de ejercicios ($RE_b$ y $RE_c$) acabando con un segundo bloque de contenidos ($BC_d$) y un último bloque de ejercicios ($RE_e$).

TABLA I. ACTIVIDADES EN LAS QUE SE HA UTILIZADO LA HERRAMIENTA SOCRATIVE, INDICADO CON UN ASTERISCO (*).

| Bloque | $BC_a$ | $RE_b$ | | $RE_c$ | | $BC_d$ | $RE_e$ | |
|---|---|---|---|---|---|---|---|---|
| Métrica | $P_a$ | $R_b$ | $D_b$ | $R_c$ | $D_c$ | $P_d$ | $R_e$ | $D_e$ |
| Socrative | | | | * | | * | * | * |

*C. Mediciones*

El objetivo del experimento era ver si la participación de los alumnos aumentaba cuando se utilizaba la herramienta *Socrative* en las clases. Para ello, se midió el número de alumnos participantes comparando los casos en los que se utilizaba la herramienta y en los que no se utilizó.

En el diseño del experimento, se alternaba la utilización de la herramienta *Socrative* a lo largo de las distintas actividades de una sesión de clase. Esta alternancia se produjo para los tres tipos de actividades que hemos medido: preguntas durante un bloque de explicación de contenidos ($P_n$), ejercicios resueltos ($R_n$) y dudas sobre la resolución de ejercicios ($D_n$). En la Tabla I se muestra la temporalización de las actividades durante una sesión y las variables que se midieron a lo largo de dichas actividades. Igualmente, se ha indicado con un asterisco (*) aquellas en las que se utilizó *Socrative* para interactuar con los alumnos. Para un mejor entendimiento, añadiremos un asterisco en el subíndice de todas aquellas mediciones que se realizaron en una actividad en las que se utilizó *Socrative*, como por ejemplo, $R_{c*}$.

IV. RESULTADOS

En la Tabla II, se muestran los resultados de cada medición de las diferentes sesiones. Es necesario indicar, que para calcular los porcentajes de participación el número total de alumnos no es siempre el mismo. Para aquellas actividades en las que los alumnos participaban mediante métodos tradicionales como levantar la mano para preguntar, el porcentaje se calculó respecto al total de la clase (n). Para los casos en los que se utilizó la herramienta Socrative, el porcentaje se calculó en base a los alumnos que tenían teléfono móvil u ordenador para utilizar la herramienta ($n_S$), ya que nos encontramos en la situación de que no todos los alumnos de la clase tenían acceso a un dispositivo móvil o a un ordenador portátil durante el desarrollo de la clase.

TABLA II. NÚMERO DE ALUMNOS QUE HAN PARTICIPADO DURANTE LAS ACTIVIDADES Y PORCENTAJE RESPECTO AL TOTAL DE ALUMNOS.

| | $A_1$ | | $A_2$ | | $A_3$ | | $A_4$ | |
|---|---|---|---|---|---|---|---|---|
| | $x_i$ | % | $x_i$ | % | $x_i$ | % | $x_i$ | % |
| n | 36 | | 66 | | 62 | | 28 | |
| $n_S$ | 34 | | 56 | | 58 | | 26 | |
| $P_a$ | 6 | 16,7 | 4 | 6,1 | 4 | 6,5 | 8 | 28,6 |
| $R_b$ | 26 | 72,2 | 52 | 78,8 | 46 | 74,2 | 20 | 71,4 |
| $D_b$ | 4 | 11,1 | 4 | 6,1 | 6 | 9,7 | 6 | 21,4 |
| $R_{c*}$ | 34 | 100 | 48 | 85,7 | 58 | 100 | 26 | 100 |
| $D_c$ | 0 | 0,0 | 0 | 0,00 | 2 | 3,2 | 4 | 14,3 |
| $P_{d*}$ | 12 | 35,3 | 12 | 21,4 | 8 | 13,8 | 12 | 46,2 |
| $R_{e*}$ | 34 | 100 | 56 | 100 | 58 | 100 | 26 | 100 |
| $D_{e*}$ | 4 | 11,8 | 8 | 14,3 | 2 | 3,5 | 4 | 15,4 |

Una vez se obtuvieron los resultados de participación de los alumnos, que se muestran en la Tabla II, se analizó si dichos resultados mostraban que la utilización de los dispositivos móviles, y en particular de la aplicación *Socrative*, aumentaba la participación de los alumnos. Con este objetivo, se realizó un estudio cuantitativo de las variables medidas, para comprobar si existían o no diferencias significativas entre los valores observados cuando se utilizaban los móviles y cuando no.

El problema que tratamos de estudiar es el típico problema en el que se evalúa un determinada variable de una población inicialmente, se aplica un cambio en dicha población y posteriormente se vuelve a medir la variable estudiada. En nuestro caso, el cambio al que se sometió la población fue la utilización de pulsadores virtuales, en concreto, de la aplicación *Socrative*. La variable que se estudió antes y después del *tratamiento* fue la participación de los alumnos. En una situación ideal, deberíamos estudiar a cada individuo de forma independiente y ver si participaban en las clases antes y después, mediante la consulta de dudas, entrega de ejercicios, etc. En cualquier caso, una de las premisas del estudio era que los alumnos participaran de forma anónima, y esta característica podía ser ofrecida por la herramienta

*Socrative*. Por este motivo, no se llevó a cabo este análisis individualizado, por lo que trabajamos con porcentajes globales de población que participaba antes y después de utilizar la herramienta.

Por lo tanto, estamos ante un caso con una única variable independiente de tipo nominal y con dos posible valores: utilizar o no *Socrative*. La variable dependiente, era el porcentaje de participación de los alumnos de la clase, es decir, una variable de razón. Finalmente no se pudo garantizar la independencia de los alumnos entre los ya que podían estar solapados. Bajo estas condiciones, el análisis estadístico que se realizó fue el test *t* independiente para muestras apareadas (paired t-test).

Se comparó de forma independiente para cada tipo de actividad, si existía o no una diferencia significativa entre las muestras en los casos en los que se utilizó *Socrative* y en los casos en los que no. Por ejemplo, se comparó la variable correspondiente al porcentaje de alumnos que preguntaron durante un bloque de exposición de contenidos ($P_n$), para los casos en los que se utilizó *Socrative* ($P_a$) y en el que no ($P_d$). Esto se repitió de la misma forma para las otras dos variables: respuestas a ejercicios ($R_n$) y dudas durante la resolución de problemas ($D_n$). De igual forma, se comprobó que no existían diferencias significativas entre los casos en los que las variables de participación de una misma actividad utilizaban el mismo método, por ejemplo, no debían existir diferencias significativas entre $D_b$ y $D_c$ porque en ninguna de las dos se utilizó la herramienta *Socrative*. La Tabla III muestra las parejas de variables que se compararon mediante el test estadístico seleccionado.

TABLA III. RESULTADOS DEL P-VALOR PARA LOS DISTINTOS T-PAIRED TESTS CON UN INTERVALO DE CONFIANZA DEL 90%.

| | $P_a$ | $R_b$ | $D_b$ | $R_{c*}$ | $D_c$ | $P_{d*}$ | $R_{e*}$ | $D_{e*}$ |
|---|---|---|---|---|---|---|---|---|
| $P_a$ | | | | | | 0,01 | | |
| $R_b$ | | | | 0,02 | | | 0,001 | |
| $D_b$ | | | | | 0,01 | | | 0,77 |
| $R_{c*}$ | | | | | | | 0,39 | |
| $D_c$ | | | | | | | | 0,17 |
| $P_{d*}$ | | | | | | | | |
| $R_{e*}$ | | | | | | | | |
| $D_{e*}$ | | | | | | | | |

De esta forma, la hipótesis nula del estudio ($H_0$) era que no existía una diferencia en la participación de los alumnos entre los dos casos analizados por el test. La hipótesis alternativa ($H_1$) indicaba que sí que existía diferencia entre ambos casos. Los casos en el que el ρ-valor fue menor que 0,05 rechazamos la hipótesis nula y aceptamos la hipótesis alternativa.

En la Tabla III podemos ver los ρ-valor obtenidos en el test mediante la herramienta de análisis estadístico R, con el test indicado, y con un intervalo de confianza del 90%.

## V. Discusión

La parte de discusión de resultados se divide en dos bloques fundamentales. Por un lado trataremos el análisis y discusión de los resultados obtenidos en las métricas utilizadas sobre la participación de los alumnos. En una segunda fase analizaremos la experiencia de los profesores a partir de comentarios e impresiones de la utilización de la herramienta. Esta parte de discusión es totalmente subjetiva pero la consideramos igual de importante dentro de las aportaciones de este artículo.

### A. Resultados

El estudio estaba diseñado para mostrar que *Socrative* modificaba la participación de los alumnos en clase. Por tanto, si esto se cumplía debíamos de ver que en aquellos test en los que comparamos dos casos en los que en uno se utilizó *Socrative* y en otro no, obtuvieron un ρ-valor inferior a 0,05, ya que esto indicaría que hay diferencias significativas entre ambos casos. Por el contrario, aquellos casos en los que se utilizó *Socrative* en los dos, o bien en ninguno de ellos, el valor ρ-valor debería de ser mayor que 0,05 ya que no se deberían mostrar diferencias significativas entre ambos casos.

El análisis de los resultados lo vamos a dividir en tres partes, correspondientes a cada uno de los tipos de participaciones o actividades que realizaban los alumnos. El primer análisis corresponde al caso de preguntas realizadas por los alumnos durante un bloque de explicación de contenidos ($P_n$). Para este caso se tomaron dos medidas, una en el caso de utilizar *Socrative* ($P_d$) y otra en el que no ($P_a$). El ρ-valor obtenido en el test es de 0,01, y por tanto podemos rechazar la hipótesis nula, y afirmar que la participación del alumnado varía cuando se utiliza *Socrative*. Podemos concluir entonces que existen evidencias de que la utilización de dispositivos móviles durante las explicaciones de teoría del profesor, puede favorecer la participación de los alumnos y aumentar la interacción entre profesor y alumno, rompiendo las barreras culturales que a veces se crean entre ellos, especialmente en clases de explicación magistral.

El segundo caso a analizar es el caso de la resolución de ejercicios ($R_n$). Para este caso, vimos que los ρ-valor entre el caso en que no se utilizó *Socrative* y los dos casos en que sí que se utilizó son 0,02 y 0,001. Por lo tanto, podemos afirmar que existen diferencias significativas ya que podemos rechazar la hipótesis nula. Además, si analizamos las diferencias entre los dos casos en que sí que se utilizó *Socrative*, podemos ver que no hay diferencias significativas ya que el ρ-valor es de 0,39. De esta forma podemos concluir de nuevo que el uso de pulsadores virtuales, como es *Socrative*, aumenta la participación de los alumnos también en el caso de realizar actividades durante una sesión presencial y compartir los resultados de sus cálculos/desarrollos con el resto de la clase. Vemos como de nuevo estas herramientas ayudan a que el alumno no tenga vergüenza a compartir sus resultados durante la clase, probablemente por el hecho de poder hacerlo de forma anónima.

Sin embargo, todo cambia para el último caso en el que se han estudiado las preguntas que realizan los alumnos cuando un profesor resuelve una actividad en clase que previamente los alumnos hayan intentado resolver durante la misma clase ($P_n$). En estos casos no se pudo apreciar una diferencia significativa entre los casos en los que se ha utilizado *Socrative* y los que no. Concluimos de esta forma que los alumnos llevan a cabo el mismo número de preguntas cuando se trata de resolver dudas de la resolución de un ejercicio, independientemente de que dispongan de *Socrative* o no. Esto nos hace pensar que en este tipo de actividad, se sienten menos cohibidos a participar, y que el hecho de utilizar una herramienta que les permita hacerlo de forma anónima, no proporciona un valor añadido.

*B. Experiencia Docente*

La primera preocupación con la que nos encontramos ante la utilización de este tipo de herramientas fue con el hecho de excluir a una serie de alumnos por no disponer de móvil o portátil. Analizando los números vimos que el porcentaje de alumnos que no disponen de estas herramientas es muy bajo, de hecho, menor de lo que esperábamos. Una práctica positiva para el futuro sería comprobar previamente que todos los alumnos tienen acceso de una u otra forma a la aplicación y, si esto no es así, trasladar la clase a un aula con ordenadores donde todos los alumnos puedan participar. En cualquier caso, hay que conseguir que ningún alumno quede excluido ya que de no ser así se conseguiría el efecto contrario del buscado con esta herramienta.

Un segundo tema a comentar son algunos problemas técnicos sufridos durante la utilización de la herramienta. Uno de los profesores tuvo problemas con la aplicación ya que ésta se desconectaba continuamente y, por tanto, a los alumnos no les aparecía la actividad. Podemos pensar que estos problemas se debían a algún tipo de fallo de conexión de la aplicación del profesor al utilizarla desde un móvil con conexión 3G. Los otros profesores utilizaron un ordenador portátil con conexión wifi y no experimentaron ningún problema.

Un hecho positivo y que nos sorprendió es que los alumnos se adaptaron muy rápidamente a la utilización de la herramienta. Inicialmente pensábamos que la dinámica de utilizarla herramienta sería lenta pero los alumnos fueron capaces de utilizarla prácticamente sin ningún tipo de explicación. Así que el tiempo utilizado para explicar el funcionamiento de la aplicación fue mínimo ya que ésta resultó ser de una alta usabilidad y muy intuitiva.

VI. CONCLUSIÓN

En este trabajo hemos mostrado, con evidencias, que la utilización de dispositivos móviles facilita a los alumnos la participación en las clases presenciales y que, por tanto, su contribución en las mismas también aumenta. Para ello diseñamos un experimento compuesto de varias fases y ejercicios en los que se fue alternando la utilización de la herramienta estudiada con los métodos tradicionales de participación de los alumnos, es decir, levantar la mano y hacer una intervención de viva voz. En el experimento, se midió una serie de métricas para evaluar la participación de los alumnos a la hora de realizar preguntas, resolver ejercicios y tenerlos resueltos de forma correcta. Finalmente, el experimento diseñado se aplicó sobre cuatro grupos distintos de alumnos.

Pudimos comprobar que existen evidencias de que la utilización de dispositivos móviles, aumenta el número de preguntas que hacen los alumnos al profesor durante la realización de la clase.

Las evidencias encontradas nos hacen pensar que existe una relación causa-efecto y que este tipo de herramientas podría ofrecer más ventajas a los profesores para conseguir mitigar los inconvenientes de la no participación de los alumnos en clase. Como trabajo futuro proponemos estudiar de forma más profunda este tipo de herramientas mediante el aumento de la muestra utilizada y la mejora de la estratificación de las muestras.

Como segundo grupo de trabajos futuros incluimos el hecho de intentar integrar los dispositivos móviles también en la realización de otros tipos de prácticas, como por ejemplo, la co-evaluación entre iguales. Algunos de los compañeros de nuestra universidad ya han llevado a cabo algún tipo de experiencia con las que han quedado muy satisfechos [9, 10].

Como punto final comentar alguna de las ideas que nos han surgido durante la realización de las actividades. Una de las que nos parece más interesante es la de compartir con los alumnos el visionado de la aplicación del profesor con un proyector. En esta parte de la aplicación aparecen las respuestas que van dando los alumnos de forma interactiva e inmediata mediante el uso de gráficos, listados, etc. Esto podría servir para llevar a cabo juegos entre los alumnos a la hora de realizar actividades, motivando al alumnado a dar una respuesta rápida y correcta de los ejercicios planteados. Ha quedado demostrado en muchos estudios que la utilización de juegos mejora el aprendizaje de los alumnos [5, 11].

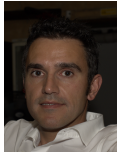
**Carlos Guerrero** obtuvo el grado de Doctor en Informática en la Universitat de les Illes Balears en el año 2012. Actualmente ocupa una plaza de profesor colaborador en el área de Arquitectura y Tecnologías de los Computadores en el Departamento de Matemáticas e Informática de la misma universidad. Sus temas de investigación están relacionados con el rendimiento web, la ingeniería web, las aplicaciones web y móviles, la minería de datos, la computación en la nube, la innovación en la educación y la utilización de las tecnologías de la información en la educación. Es autor de 12 artículos publicados en congresos internacionales y en revistas. Igualmente ha formado parte de un gran número de comités de programa y de revisión. Ha participado en proyectos de investigación nacionales e internacionales y ha liderado proyectos locales de innovación docente durante los últimos cinco años.

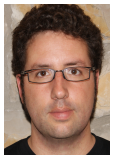
**Antoni Jaume-i-Capó** obtuvo el doctorado en Informática por la Universitat de les Illes Balears en 2009. Desde 2005 trabaja en el Departamento de Ciencias Matemáticas e Informática de la Universidad de las Islas Baleares. Sus principales intereses de investigación incluyen la visión computacional, las interfaces basadas en la visión, serious games, rehabilitación, co-evaluación y motivación. Ha publicado en diversas revistas internacionales su trabajo de investigación. También ha participado y liderado proyectos de investigación financiados por diferentes organismos (Gobierno Regional, Gobierno Español, Unión Europea).

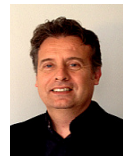
**Carlos Juiz,** Doctor en Informática en 2001 por la Universitat de les Illes Balears (UIB), España. Después de varios cometidos en la industria de las Tecnologías de la Información (TI) durante 14 años, en la actualidad ejerce como Profesor Titular de Universidad en la UIB. Ha sido investigador invitado en la Universidad de Viena (2003) y profesor invitado en Stanford University (2011). Ha escrito más de 150 artículos en revistas, actas de congresos, revisiones y capítulos de libros. Es colaborador habitual en asignaturas de máster y doctorado nacionales e internacionales. Ha desarrollado proyectos europeos de investigación y desarrollo, así como de docencia y gestión universitaria. También ha dirigido proyectos de cooperación internacional principalmente en Túnez, Marruecos, Paraguay y Ecuador. Ha desempeñado diversos cargos de gestión universitaria, entre los que se destacan Director de Prospectiva y Vicerrector de Tecnologías de la Información (TI) en la UIB. Actualmente es miembro del consejo de dirección de varios clústeres empresariales relacionados con la industria del e-Turismo y las TI. Ha sido conferenciante invitado en múltiples eventos de ámbito nacional e internacional. Miembro Sénior de IEEE, Miembro Sénior de ACM, miembro del Dominio de Cloud Computing de la IFIP y experto invitado en la ITU. Es líder del subgrupo de Gobierno de las TI en AENOR. Sus temas de interés son la evaluación del rendimiento de los sistemas informáticos, la gestión y gobernanza de las TI y la estrategia e innovación del e-Turismo.

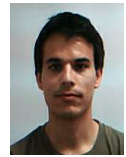
**Isaac Lera,** Doctor en Informática por la Universitat de les Illes Balears en 2012. Desde 2006 ha estado vinculado con el Departamento de Ciencias Matemáticas e Informática en la misma universidad. Actualmente imparte clases en el grado de Informática y en el master universitario en tecnologías de la información. Sus líneas de interés son: web semántica, integración de grandes volúmenes de datos, datos abiertos, arquitectura y rendimiento de sistemas informáticos. Posee múltiples publicaciones en revistas y congresos y ha participado en diferentes proyectos nacionales e internacionales, algunos de los cuales han sido de innovación docente.